# THERMAL CONDUCTANCE OF A WEAKLY COUPLED QUANTUM DOT

Margarita Tsaousidou[*]

*Materials Science Department, University of Patras, Patras 26 504, Greece*

George P. Triberis

*Physics Department, University of Athens, Panepistimiopolis, 157 84, Zografos, Athens, Greece*

## Abstract

We calculate the electronic contribution to the thermal conductance, κ, in a quantum dot that is weakly coupled via tunnel barriers to two electrons reservoirs. A linear response model is used for the calculation of the heat current Q through the quantum dot when a small temperature difference ΔT and a small voltage difference ΔV are applied between the two reservoirs. We find that κ oscillates as a function of the Fermi energy in the reservoirs. The periodicity of these oscillations is the same as this of the Coulomb blockade oscillations of the conductance. In the quantum limit the peak values of the thermal conductance tend rapidly to zero as exp(-ΔE/$k_B$T) where ΔE is the spacing between the energy levels in the dot. This surprising result implies the failure of the Wiedemann-Franz law in the quantum limit and it could lead to structures with large figure of merit ZT. We have also examined the behavior of κ in the classical limit and we find that the Wiedemann-Franz is recovered.

Since the late 1980's the study of the conductance, G, and the thermopower, S, of quantum dots (QD's) in the Coulomb blockade regime has attracted significant theoretical and experimental interest. The study, however, of the electron thermal conductance in this regime has received much less attention [1]. The purpose of the present work is to provide the first theoretical formalism [2] for the calculation of the electronic contribution to the thermal conductance, κ, of a weakly QD in the Coulomb blockade regime. Our analysis is based on a master equation approach [3,4]. Here we neglect effects due to the virtual tunneling (cotunneling) of electrons through the dot. By using a linear response model we calculate the heat current Q through the dot in the presence of a small voltage difference ΔV and a small temperature difference ΔT between the reservoirs.

We consider a quantum dot with equidistant energy levels $E_m$ ($m = 1,2,3,...$) which is weakly coupled to two electron reservoirs. Each energy level in the dot contains 0 or 1 electrons. The charging energy $E_c = e^2 / 2C$, where C is the capacitance between the dot and the reservoirs, is larger than the level spacing ΔE. The tunnel rate from level m to the left (right) reservoir is denoted as $\Gamma_m^L$ ($\Gamma_m^R$). The electrostatic energy for a dot with N electrons is $U(N) = (Ne)^2 / C - Ne\phi_{ext}$ where $\phi_{ext}$ is the electrostatic potential from external charges. In equilibrium both reservoirs have the same temperature T and the same chemical potential $E_F$. The energy distribution in the reservoirs is the Fermi Dirac function $f(E - E_F) = 1/\{1 + \exp[(E - E_F)/k_B T]\}$.

We assume that the temperature and the chemical potential in the right reservoir are raised by $\Delta T$ and $e\Delta V$, respectively. In the regime of linear response the heat flow is related to $\Delta T$ and $\Delta V$ by the equation

Q = M$\Delta$V + K$\Delta$T     (1)

where M and K are transport coefficients. The scope of our work is to calculate Q and determine M and K. Then, the thermal conductance, $\kappa$, can be readily obtained.

Assuming that $\Delta T$ and $\Delta V$ are small we find, after some complicated algebra, that the stationary heat current through the dot is given by

$$Q = -\sum_{m=1}^{\infty}\sum_{N=1}^{\infty} \frac{\Gamma_m^L \Gamma_m^R}{\Gamma_m^L + \Gamma_m^R} P_{eq}(N) P_{eq}(E_m \mid N)[1 - f(\varepsilon - E_F)]\left[(\varepsilon - E_F)\frac{e\Delta V}{k_B T} + (\varepsilon - E_F)^2 \frac{\Delta T}{k_B T^2}\right]$$
     (2)

where, $\varepsilon = E_m + U(N) - U(N - 1)$. $P_{eq}(N)$ is the probability that the dot has N electrons in equilibrium and $P_{eq}(E_P \mid N)$ is the conditional probability that in equilibrium the m state is occupied given that the dot has N electrons.

The transport coefficients M and K are readily calculated by using Eqs.(1) and (2). M is related to the thermopower $S$ by Onsager's relation [5,6]: S=M/TG. Thus, the expression for the thermpower is

$$S = -\frac{e}{k_B T^2 G}\sum_{m=1}^{\infty}\sum_{N=1}^{\infty} \frac{\Gamma_m^L \Gamma_m^R}{\Gamma_m^L + \Gamma_m^R} P_{eq}(N) P_{eq}(E_m \mid N)[1 - f(\varepsilon - E_F)](\varepsilon - E_F)$$
     (3)

where, G is the conductance in quantum dot given by the expression [3]:

$$G = \frac{e^2}{k_B T}\sum_{m=1}^{\infty}\sum_{N=1}^{\infty} \frac{\Gamma_m^L \Gamma_m^R}{\Gamma_m^L + \Gamma_m^R} P_{eq}(N) P_{eq}(E_m \mid N)[1 - f(\varepsilon - E_F)]$$
     (4)

Moreover, we find that the thermal coefficient K is

$$K = -\frac{1}{k_B T^2}\sum_{m=1}^{\infty}\sum_{N} \frac{\Gamma_m^L \Gamma_m^R}{\Gamma_m^L + \Gamma_m^R} P_{eq}(N) P_{eq}(E_m \mid N)[1 - f(\varepsilon - E_F)](\varepsilon - E_F)^2$$
     (5)

Finally, the thermal conductance, $\kappa$, is obtained by the standard relation [6]:

$$\kappa = -K - S^2 GT$$
     (6)

We should mention here that the expression (3) we derived for S is identical to the expression (3.11) obtained by Beenakker and Staring [3] who calculated the electron current through the dot instead of the heat current. The approach we use here for the calculation of S (the so-called $\Pi$-approach [7]) has been applied before for the calculation of the phonon-drag magnetothermopower in GaAs/AlGaAs heterojunctions [8] and the phonon-drag thermopower in ballistic quantum wires [9]. The advantage of our approach is that we can calculate also the thermal coefficient K and by using Eq.(6) the thermal conductance $\kappa$.

In Fig. 1 we show the thermal coefficient K for a three level dot in the quantum regime where $\Delta E >> k_B T$. K is computed by using Eq. (5). The energy level separation is taken to be $\Delta E = 0.5 e^2/2C$. $L_0$ is the Lorentz number and $G_0 = (e^2/k_B T)\Gamma^L \Gamma^R (\Gamma^L + \Gamma^R)^{-1}$. We see that K shows a double peak structure with the same periodicity as that of the Coulomb-blockade oscillations of the conductance. The effect of temperature on this structure is shown in Fig. 1b. In Fig.1 we also show the calculation of $S^2 GT$ by using Eqs. (3) and (4). We note that for clarity reasons the results are divided by $L_0 TG_0$. The two curves are indistinguishable. Inspection of Eq.(6) shows that *the electronic contribution to the thermal conductance goes to zero*. The above result has also been verified analytically. Namely, in the quantum regime the main contribution to the summations over m and N is made when $m = N = N_{min}$ [4] where $N_{min}$ corresponds to the number of electrons for which the absolute value for $\Delta^{qu}(N) = E_N + U(N) - U(N-1) - E_F$ is minimum. We define $\Delta^{qu}(N_{min}) \equiv \Delta^{qu}_{min}$. When $T \to 0$ the probability distributions are written $P_{eq}(N_{min}) \approx [1 + \exp(\Delta^{qu}_{min}/kT)]^{-1}$ and $P_{eq}(E_P | N) \approx 1$ [4]. Now, the expressions for G, S, K, are considerably simplified:

$$G = -e^2 \frac{\Gamma^L \Gamma^R}{\Gamma^L + \Gamma^R} f'(\Delta^{qu}_{min}) \ [4] \tag{7}$$

$$S = -\frac{1}{eT} \Delta^{qu}_{min} \tag{8}$$

$$K = \frac{1}{T} \frac{\Gamma^L \Gamma^R}{\Gamma^L + \Gamma^R} (\Delta^{qu}_{min})^2 f'(\Delta^{qu}_{min}). \tag{9}$$

Inspection of the above equations shows that $K = -S^2 GT$ and consequently *the thermal conductance tends to zero* when $\Delta E >> k_B T$. We should remark, that the approximated expression (8) for the thermopower does not reveal the fine structure of S of the thermopower oscillations [3]. However, it is adequate for the purposes of our analysis here which concerns the fact that the peak values of $\kappa$ are found to be very small in comparison with what we would expect if the Wiedemann-Franz law remained valid in the quantum limit.

In order to reveal the analytical expression that shows the exact way that $\kappa$ tends to zero at low temperatures we explore the non-zero terms in $\kappa$. It can be shown [10] that the summation over $m = N_{min} + 1$ και $N_{min} - 1$ in Eqs.(3)-(5) produces corrections of the order of $\exp(-\Delta E/k_B T)$. The analytical expression for $\kappa$ derived when first order terms in $\exp(-\Delta E/k_B T)$ are kept in Eq. (6) is

$$\kappa = k_B \frac{\Gamma^L \Gamma^R}{\Gamma^L + \Gamma^R} \frac{(1 + e^{\Delta E/k_B T}) e^{\Delta^{qu}_{min}/k_B T}}{e^{\Delta E/k_B T}(1 + e^{2\Delta^{qu}_{min}/k_B T}) + e^{\Delta^{qu}_{min}/k_B T}(1 + e^{2\Delta E/k_B T})} \left(\frac{\Delta E}{k_B T}\right)^2 \tag{10a}$$

and around $\Delta_{min}^{qu} = 0$ $\kappa$ is written in a very simple form

$$\kappa = k_B \frac{\Gamma^L \Gamma^R}{\Gamma^L + \Gamma^R} \frac{1}{[1 + \exp(\Delta E / k_B T)]} \left(\frac{\Delta E}{k_B T}\right)^2 \approx k_B \frac{\Gamma^L \Gamma^R}{\Gamma^L + \Gamma^R} \left(\frac{\Delta E}{k_B T}\right)^2 \exp(-\Delta E / k_B T) \qquad (10b).$$

The expression (10) for $\kappa$ in the quantum limit is in excellent agreement with the numerical results. Numerical calculations of the thermal conductance as a function $E_F$ for several values of $\Delta E/k_B T$ are shown in Fig. 2 for a dot with six electrons. The periodicity of the oscillations is $\Delta E_F = \Delta E + (e^2/C)$ which is the same as the periodicity of the Coulomb-blockade oscillations in the conductance [4]. Similar oscillations have been found in the strong tunnelling regime by Moskalets [1]. We should remark here that the first and the last peaks are not described by Eq.(10a). The expressions that described accurately the shape of the peaks of $\kappa$ when $N_{min}$ is equal either to one or to the total number of electrons in the dot are given in the full article [10].

Inspection of Eqs.(7) and (10) shows that the Wiedemann-Franz law is not valid in the quantum limit. It is found that the peak values of the Coulomb blockade oscillations of the thermal conductance are smaller than what the Wiedemann-Franz law predicts by a factor of the order of $\exp(-\Delta E/k_B T)(\Delta E/k_B T)^2$. Another interesting feature of $\kappa$ is that remains almost constant in the region around $\Delta_{min}^{qu} = 0$ when $\Delta E/k_B T >> 1$. This fact gives the possibility of achieving very large values for the figure of merit in structures where the phonon contribution to the thermal conductance is very small. This could make the structure we study a very promising candidate for improved thermoelectric materials. [Detailed discussion is given in Ref. [10].]

In Fig. 3(a) we present the calculated figure of merit ZT for $\Delta E/k_B T$ =12.5 and 8.3. (The phonon contribution is neglected.) The existence of two local maxima are associated to the maximization of $S^2 GT$ at $\Delta_{min}^q / k_B T \approx 2.4$ and $-2.4$. The maximum value of ZT depends exponentially on the ratio $\Delta E/k_B T$ according to the relation [10]

$$ZT^{max} = 0.44 \, e^{\Delta E / k_B T} \left(\frac{k_B T}{\Delta E}\right)^2 \qquad (11)$$

In Fig. 3(b) we also depict the calculated ZT for $\Delta E/k_B T$=5.

We should stress here that the contribution of phonons could have as a result the strong suppression of ZT. However, the fact that the electronic contribution to the thermal conductance tends to zero for large values of $\Delta E/k_B T$ could be of great importance for designing structures where the phonon contribution to the thermal conductance can be minimized.

In the classical limit where $\Delta E << k_B T << e^2/2C$ it can be shown that the thermal conductance is given by [10]

$$\kappa = L_0 T G \left[ 1 + \frac{1}{4\pi^2} \left( \frac{\Delta(N_{min})}{k_B T} \right)^2 \right] = L_0 T G \left( 1 + \frac{S^2}{3L_0} \right) \qquad (10)$$

where, $\Delta(N) = U(N) - U(N-1) + \bar{\mu} - E_F$ with $\bar{\mu}$ being the chemical potential in the dot and $N_{min}$ is the number of electrons that minimizes the $|\Delta(N)|$. For simplicity reasons we have assumed that the density of states and the tunnel rates do not depend on energy. Inspection of Eq.(10) shows that when $S^2 << L_0$ *one finds the Wiedemann-Franz relation:* $\kappa \approx L_0 T G$.

In Fig. 4 we show the thermally broadened conductance and thermal conductance peaks in the classical regime. The energy $\Delta(N_{min})$ is proportional to the Fermi energy in the reservoirs. $G_{max}$ is the peak height of the Coulomb-blockade oscillations and $\kappa_{max} = L_0 T G_{max}$. We find that in the classical limit, the oscillation peaks of the conductance and the thermal conductance follow the Wiedemann-Franz law.

In conclusion, we calculated the electronic contribution to the thermal conductance in a quantum dot weakly coupled to two electron reservoirs. Analytical expressions for $\kappa$ are given for the quantum and the classical limits. In the quantum regime we find that the Wiedemann-Franz law is violated. We also show that the peak value of $\kappa$ is exponentially small of the order of $\exp(-\Delta E / k_B T)$. This finding suggests that the figure of merit could be significantly enhanced in structures where the phonon contribution to the thermal conductance becomes small. In the classical limit we show that the Wiedemann-Franz law is recovered.

**Acknowledgements**

The author MT acknowledges X. Zianni (Dept. of Applied Sciences, Technological Educational Institution of Chalkida, Greece) for giving the motivation of this work through the 'Archimides I' programme funded by the European Commission and by the Greek Ministry of Education, O.P. 'Education' (E.P.E.A.E.K.). MT would also like to acknowledge Robin Fletcher for stimulating remarks and valuable suggestions.

[*]Electronic address: rtsaous@upatras.gr

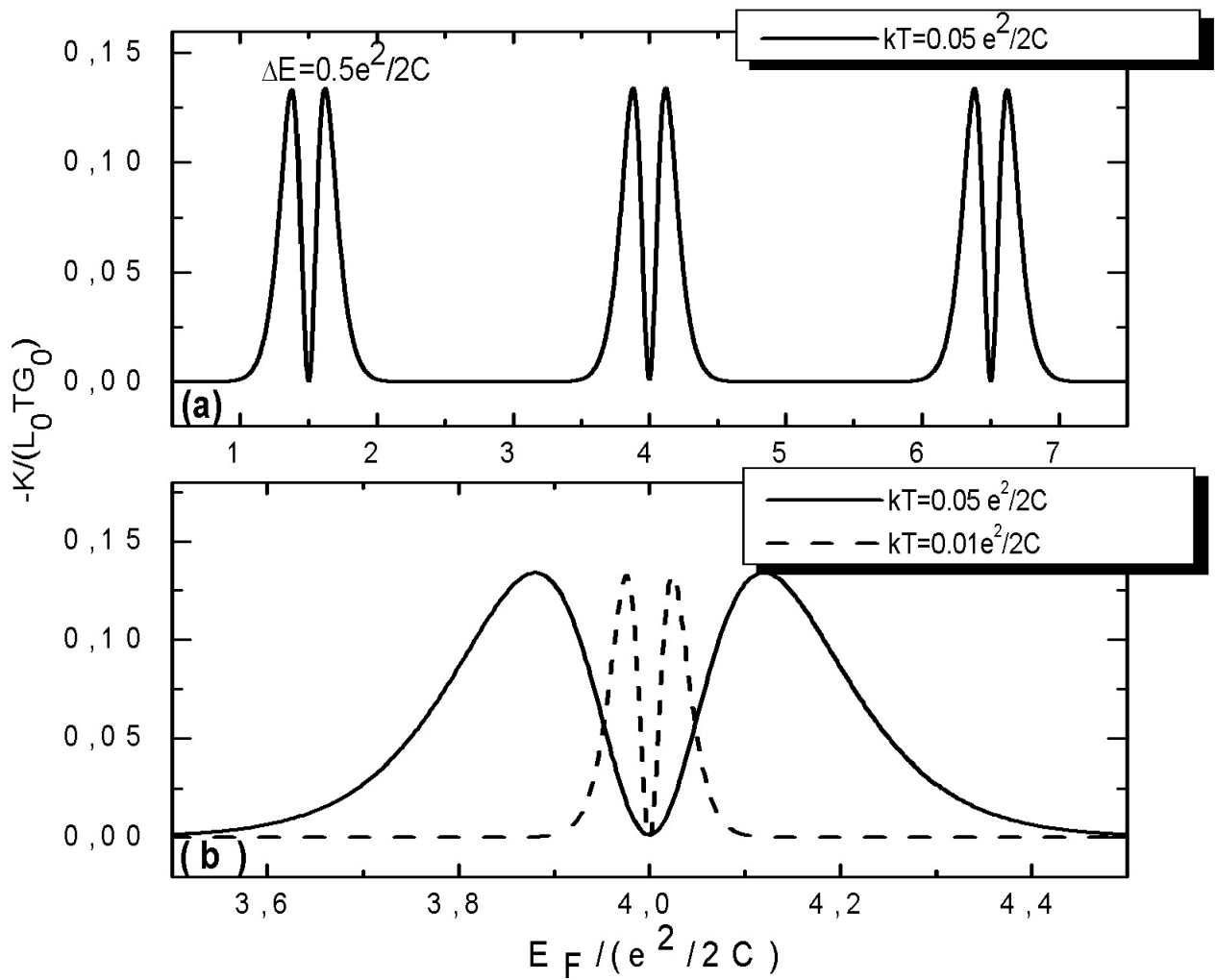

**Figure 1.** The thermal coefficient K as a function of $E_F$ in a quantum dot with three electrons. Details are given in the text.

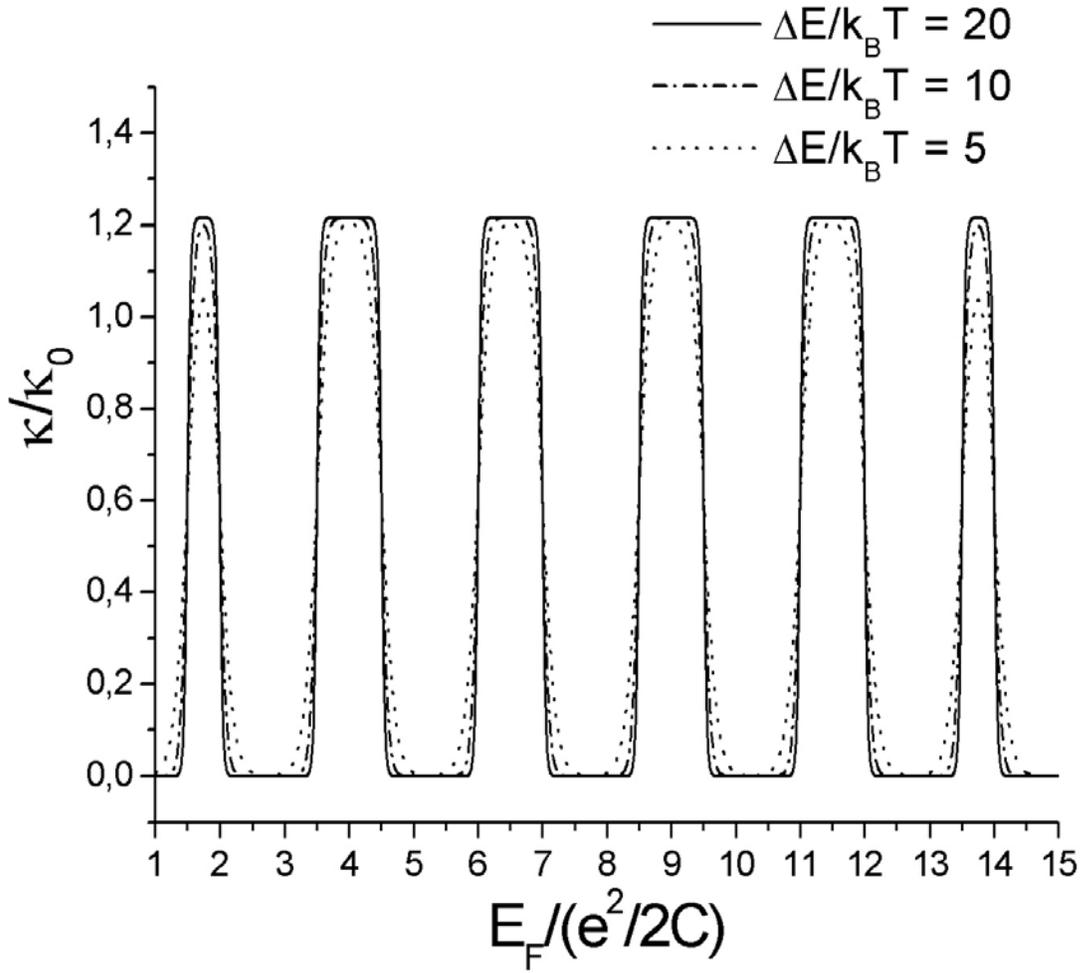

**Figure 2.** Calculations of the thermal conductance, $\kappa$, as a function of $E_F$ in a quantum dot with six electrons in the quantum limit for three values of $\Delta E/kT$. $\kappa_0$ is given by $L_0 T G_0 \exp(-\Delta E/k_B T)(\Delta E/k_B T)^2$ and $G_0 = (e^2/k_B T)\Gamma^L \Gamma^R (\Gamma^L + \Gamma^R)^{-1}$. We note that $G_0 = 4G_{max}$ where $G_{max}$ is the conductance peak around $\Delta_{min}^{qu} = 0$.

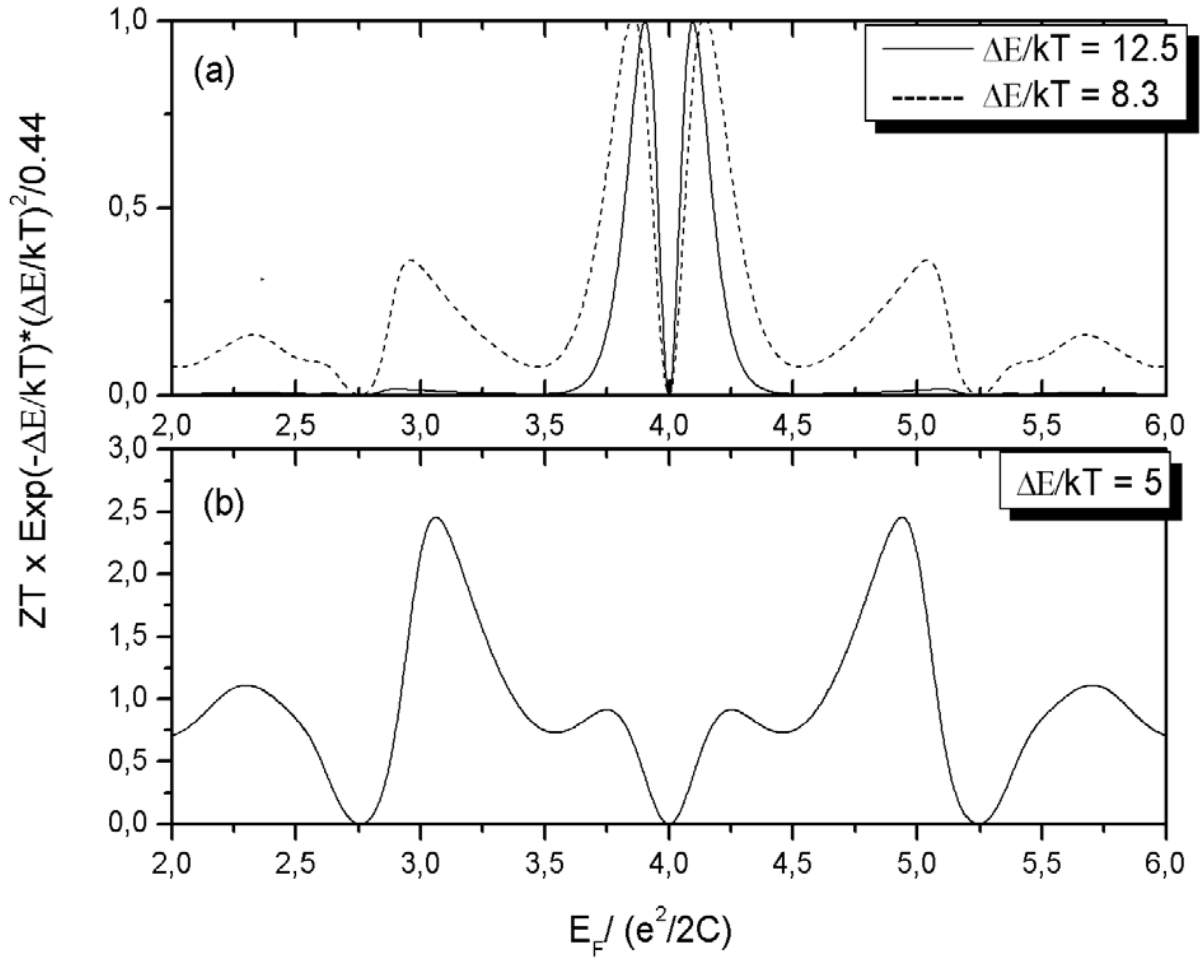

**Figure 3.** Figure of merit, ZT, as a function of $E_F$ around the region $\Delta_{min}^{qu} = 0$. We present the results for $\Delta E/k_B T$=12.5, 8.3 and 5. For clarity reasons the calculated values of ZT are multiplied by the factor $\exp(-\Delta E/k_B T)(\Delta E/k_B T)^2/0.44$.

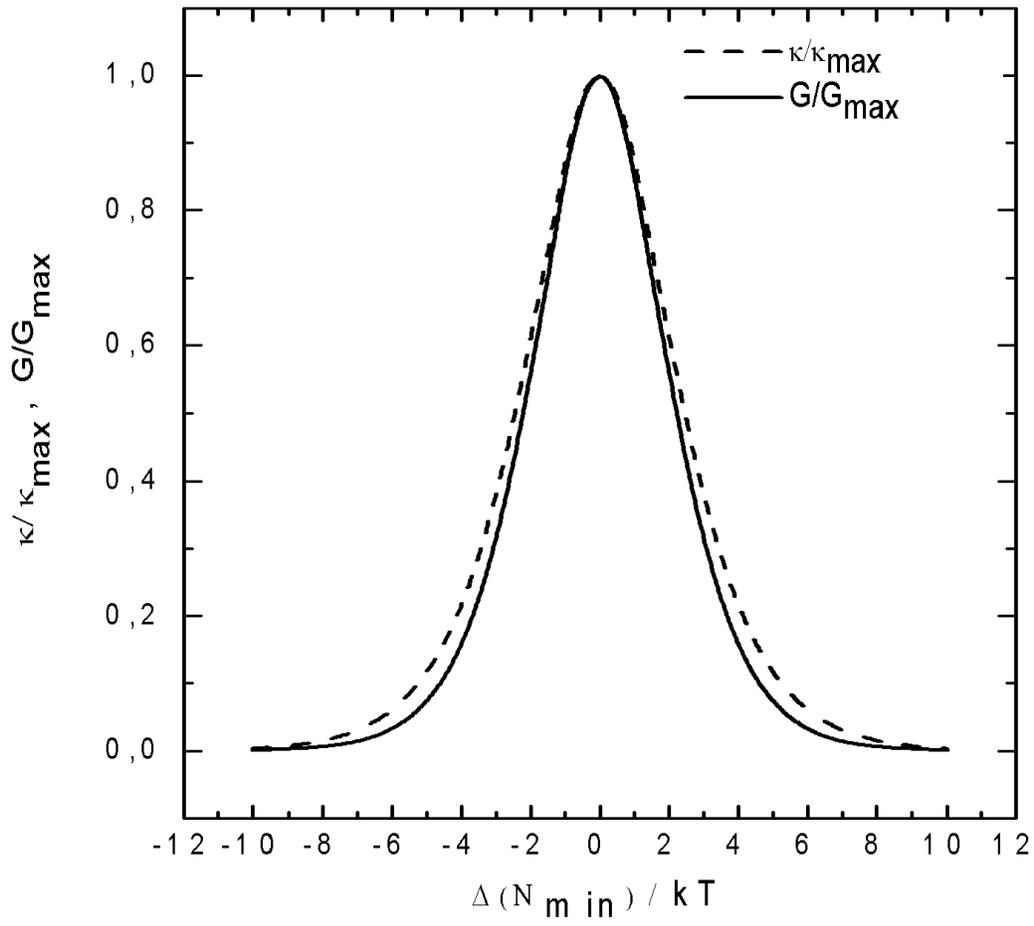

**Figure 4.** The thermally broadened conductance and thermal conductance peaks in the classical regime.